\documentclass[preprint,aps,superscriptaddress,nofootinbib,tightenlines]{revtex4}

\usepackage{epsfig}

\def\OMIT#1{{}}

\newcommand{\lsim}{\raisebox{-0.7ex}{$\stackrel{\textstyle <}{\sim}$ }}

\begin{document}

\preprint{\vbox{
\psfig{file=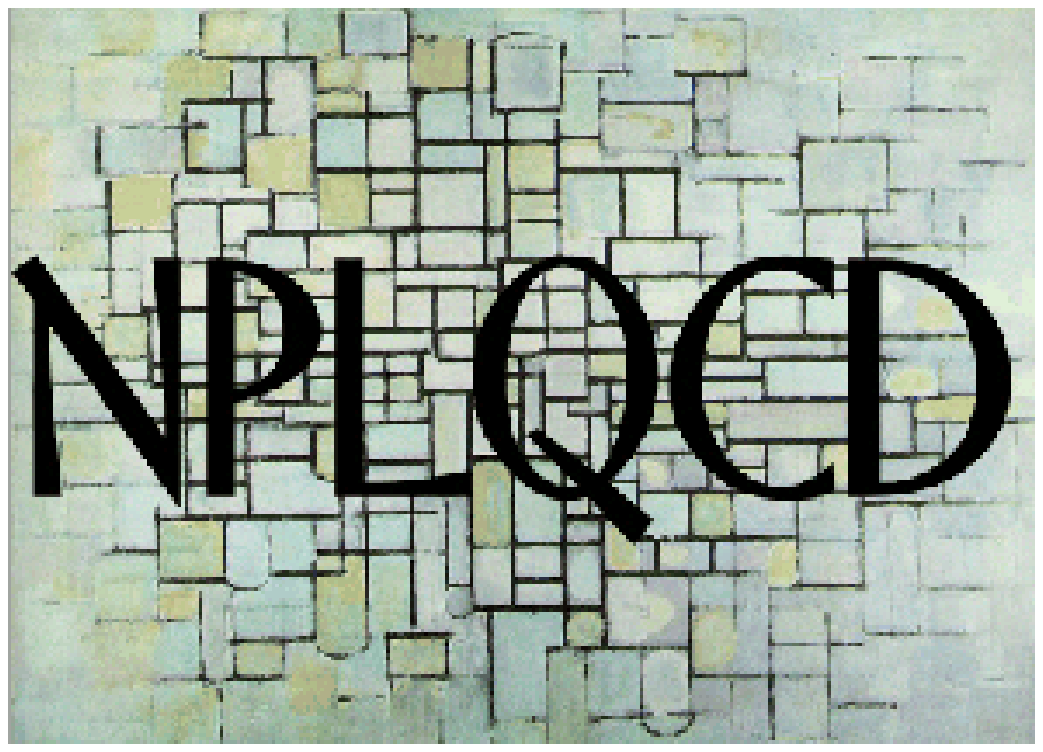,width=1.1in,angle=0}\hfill
\hbox{UNH-04-02}}}
 
\phantom{ijk}
\vskip 0.5cm
\title{In Search of the Chiral Regime}

\author{\bf Silas R.~Beane}
\affiliation{Department of Physics, University of New Hampshire,
Durham, NH 03824-3568.}
\affiliation{Jefferson Laboratory, 12000 Jefferson Avenue, 
Newport News, VA 23606.}

\vphantom{}
\vskip 0.5cm
\begin{abstract} 
\noindent A critical appraisal is given of a recent analysis of the
quark-mass  and finite-size dependence of unquenched lattice QCD data for the nucleon
mass. We use this forum to estimate the boundary of the chiral regime
for nucleon properties.
\end{abstract}

\maketitle

\noindent How low in quark masses must lattice QCD simulations of
baryon properties go in order to reach the regime where chiral
perturbation theory ($\chi$PT) is an extrapolation tool with controlled errors?
In interesting recent work~\cite{Procura,Khan:2003cu} unquenched
lattice QCD data for the nucleon mass at large pion masses, $m_\pi >
500~{\rm MeV}$, has been analyzed using baryon $\chi$PT to ${\cal O}(m_q^2)$
(without including the $\Delta$ as an explicit degree of
freedom). These papers fit the quark-mass dependence of unquenched
lattice QCD data for the nucleon mass (including the physical point)
with natural values of the strong-interaction parameters and then
successfully predict (in Ref.~\cite{Khan:2003cu}) the finite-size
dependence. These results are quite remarkable. However, little attempt is made in
Refs.~\cite{Procura,Khan:2003cu} to gauge the uncertainties associated
with the quark-mass extrapolation procedure. Given the provocative results that
are found and the rather large quark masses that are simulated, it is
essential to perform an error analysis.  We will therefore do so in
this note. And, in so doing, we will provide an estimate of the
quark-mass boundary of the chiral regime for nucleon
properties~\cite{bernard}.

\begin{figure}[ht]
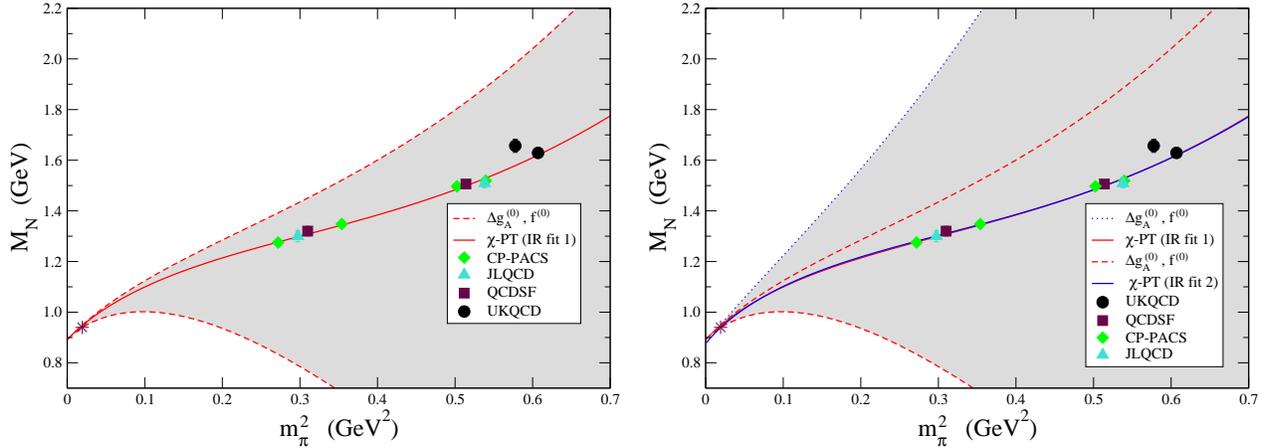

\vskip 0.4in
\centerline{{\epsfxsize=3.2in \epsfbox{swindle1.eps}}\hskip0.36cm{\epsfxsize=3.2in \epsfbox{swindle1A.eps}}} 
\caption{\it Left panel: The nucleon mass in 
baryon $\chi$PT (computed using infrared regularization) at ${\cal O}(m_q^2)$
(parameters given in Table~\ref{table:fitjunk} (Fit 1 (IR))) vs. the pion mass squared.
Symbols are unquenched lattice QCD data taken from
Ref.~\protect\cite{Khan:2003cu,UKQCD,CPPACS,JLQCD}. The solid curve is the fit curve
and the dashed curves give maximal variations due to the spread
in chiral limit values of $f$ and $g_A$, as explained in the text.
The gray region corresponds to the error associated with omitted higher orders.
Right panel: Same but with a second fit, (parameters given in Table~\ref{table:fitjunk} (Fit 2 (IR))) with
corresponding spread in the chiral limit values of $f$ and $g_A$.
(Note that there are two curves going through the data.)}
\label{fig:swindle1}
\end{figure}

The nucleon mass at ${\cal O}(m_q^2)$ in the chiral expansion, evaluated using
infrared regularization~\cite{Becher}, is given by~\cite{Procura}
\begin{eqnarray}
M_N & = & M_0 - 4 c_1 m_\pi^2-\frac{3 g_A^2}{8 \pi^2 f^2} m_\pi^3 \sqrt{1-\frac{m_\pi^2}{4 M_0^2}} \left[\frac{\pi}{2} + 
       \arctan{\frac{m_\pi^2}{\sqrt{4M_0^2 m_\pi^2- m_\pi^4}}}\right] 
 \nonumber \\ & +&   \left[e_1(\mu)+\frac{3}{32 \pi^2 f^2}
    \left( \frac{g_A^2}{M_0} + \frac{c_2}{2} \right) 
  - \frac{3}{16 \pi^2 f^2}
       \left( \frac{g_A^2}{M_0} - 8c_1 + c_2 + 4 c_3 \right)
   \ln{\frac{m_\pi}{\mu}} \right] m_\pi^4 \ .
\label{eq:IRformula}
\end{eqnarray}
Following Refs.~\cite{Procura,Khan:2003cu} we choose the parameter set
$g_A=1.267$, $f=131~{\rm MeV}$, $c_2=3.2~{\rm GeV}^{-1}$ and $c_3=-3.4~{\rm GeV}^{-1}$.  
One then easily fits the remaining low-energy constants to unquenched lattice QCD data~\cite{Khan:2003cu,UKQCD,CPPACS,JLQCD} 
over a wide range of pion
masses~\cite{Procura,Khan:2003cu}.  For instance, in
Table~\ref{table:fitjunk} (Fit 1 (IR)) we list a set of parameter
values~\footnote{We fit the full expression in eq.~(\ref{eq:IRformula}), while 
Refs.~\cite{Procura,Khan:2003cu} fit to an expanded form; hence the small differences 
in the values of the parameters.} 
which give rise to the solid curve in Fig.~\ref{fig:swindle1}
(left panel).  While one sees an
excellent description of the data with natural values of the
strong-interaction parameters (which are consistent with $\pi-N$ phase shift analyses), 
one must quantify the errors associated with omitted
higher-order effects in order to estimate the reliability of
perturbation theory over the range of fit pion masses.
\begin{table}[ht]
\caption{ Fit parameters. The scale-dependent parameter $e_1$ is evaluated at $\mu=1~\rm{GeV}$.}
\label{table:fitjunk}
\begin{tabular}{@{}c | c | c | c | c | c  }
\hline
{} & 
$f~(\rm{MeV})$ & 
$g_A$ & 
$M_0~(\rm{GeV})$ & 
$c_1~(\rm{GeV}^{-1})$ & 
$e_1~(\rm{GeV}^{-3})$ \\ 
\hline
Fit 1 (IR)& 131 & 1.267 & 0.880 & -0.95 & 2.44  \\
Fit 2 (IR)& 124 & 1.520  & 0.872 & -1.22 & 4.90  \\
Fit 1 (DR)& 131 & 1.267 & 0.885 & -0.92 & 3.61  \\
Fit 2 (DR)& 124 & 1.520  & 0.879 & -1.07 & 6.79  \\
\hline
\end{tabular}
\end{table}

There are many ways to estimate the reliability of the chiral
expansion and Refs.~\cite{Procura,Khan:2003cu} suggest one robust method.  The
difference between a matrix element at a given order in the chiral expansion evaluated with the
physical values of $g_A$ and $f$ and evaluated with their chiral-limit values
$g^{(0)}_A$ and $f^{(0)}$ is a measure of the importance of
higher-order effects. Ref.~\cite{Khan:2003cu} estimate $g^{(0)}_A\sim
1.2$ and $f^{(0)}\sim 124~{\rm MeV}$ and find little variation in
the fit curve. However, this is no surprise as these chiral-limit values
leave the ratio $g_A/f$ unchanged. Here we will consider how
variation of the chiral-limit parameters affects the chiral expansion.

It is worth considering what is known 
experimentally about $g^{(0)}_A$ and $f^{(0)}$.
At one-loop order in the chiral expansion one has the
well-known formulas~\cite{Fettes:1999wp,GandL},
\begin{eqnarray}
g_A  =  g_A^{(0)} 
\left[ 1 - { (g_A^{(0)})^2 {m_\pi^2} \over 8\pi^2 (f^{(0)})^2}
+ {4 m_\pi^2\over g_A^{(0)} } \overline{d}_{16}
\right]\ , \qquad
f =  f^{(0)} \left[ 1  + { m_\pi^2\over 8\pi^2 (f^{(0)})^2} 
\overline{l}_4  \right]
\ \ \ ,
\label{eq:SNparams}
\end{eqnarray}
where $\overline{d}_{16}$ and $\overline{l}_4$ are (scale-independent)
low-energy constants~\footnote{The barred constants contain chiral
logarithms and therefore the formulas in eq.~(\ref{eq:SNparams}) are
useful only at the physical value of the pion mass. Extrapolation formulas
are available in Refs.~\cite{Beane:2002xf,Epelbaum:2002gb}.}. 

The parameter $\overline{l}_4$ is determined from $K_{\ell 3}$ 
and $K_{\ell 2}$ decays~\cite{GandL} and from a two-loop analysis
of $\pi-\pi$ scattering~\cite{Colangelo:2001df}.  For simplicity,
we will follow Ref.~\cite{GandL} and fix $f^{(0)}= 124~{\rm MeV}$, keeping in
mind that not accounting for variation of $f^{(0)}$ will necessarily lead to an
underestimate of the final error. An analysis~\cite{Fettes:1999wp,Fettes:fd} of the process $\pi
N\rightarrow\pi\pi N$ at one-loop order in the chiral expansion
provides three distinct determinations of $\overline{d}_{16}$:
$-0.91\pm 0.74$, $-1.01\pm 0.72$ and $-1.76\pm 0.85~{\rm GeV}^{-2}$.
Taking the $1\sigma$ limits of the three determinations gives $-2.61
~{\rm GeV}^{-2} < \overline{d}_{16} < -0.17$, or
$g_A^{(0)}=1.42\pm0.10$. A recent analysis suggests that the $\Delta$
(which is not included as an explicit degree of freedom in
Refs.~\cite{Fettes:1999wp,Fettes:fd}) plays a crucial role in the
determination of $\overline{d}_{16}$ and suggests a lower central
value: $g_A^{(0)}=1.20\pm0.10$~\cite{TomWolf} .  Taking $-2.61~{\rm
GeV}^{-2} < \overline{d}_{16} < 2.43~{\rm GeV}^{-2}$ encompasses both
analyses, {\it i.e.} $\Delta g_A^{(0)}=1.10- 1.52$. As naturalness suggests
$|\overline{d}_{16}|\sim 1$, this range of values does not introduce
anomalously large low-energy constants into the chiral expansion. This
experimental/theoretical uncertainty therefore provides an estimate of
the importance of neglected higher orders in the chiral expansion~\footnote{We emphasize that
even if $\overline{d}_{16}$ (and therefore $g^{(0)}_A$) were known
with high precision, varying ${\bar d}_{16}$ over a range of natural
values would continue to be a legitimate way of estimating errors due
to omitted higher-order effects.}.  In Fig.~\ref{fig:swindle1} (left
panel), we illustrate (dashed curves) the spread of the fit curve when
one replaces $f$ and $g_A$ by their chiral limit values, including the
range of ${\bar d}_{16}$ values.

It is important to realize that all curves encompassed by the gray region of 
Fig.~\ref{fig:swindle1} (left panel), between
the dashed curves, differ only by terms that are higher order, ${\cal O}(m_q^{5/2})$, in the chiral 
expansion. The spread of curves in Fig.~\ref{fig:swindle1} (left panel) therefore 
suggests a $50\%$ error associated with neglected higher orders in the
chiral expansion at the lattice point with the lowest pion mass. 
We also perform a second fit with select chiral-limit values of
$f$ and $g_A$, see Table~\ref{table:fitjunk} (Fit 2 (IR)), and again
considering variation in the chiral limit values~\footnote{Notice that although the parameter $e_1(1~{\rm GeV})$
appears rather large, we have defined $e_1$ as in Refs.~\cite{Procura,Khan:2003cu} which has absorbed
a factor of $4$ into its definition as compared to the parameter to which naturalness arguments should be
applied~\cite{FFMS}.
Hence, all values of $e_1$ in Table~\ref{table:fitjunk} are technically quite natural.}. 
The first and second fit are shown in
Fig.~\ref{fig:swindle1} (right panel), together with the 
spread (dashed and dotted curves) of the fit curves when one replaces $f$ and $g_A$ by their
chiral limit values, including the range of ${\bar d}_{16}$ values.
The spread of curves in Fig.~\ref{fig:swindle1} (right panel), 
suggests an $80\%$ error at the lattice point with the lowest pion mass. 
\begin{figure}[ht]
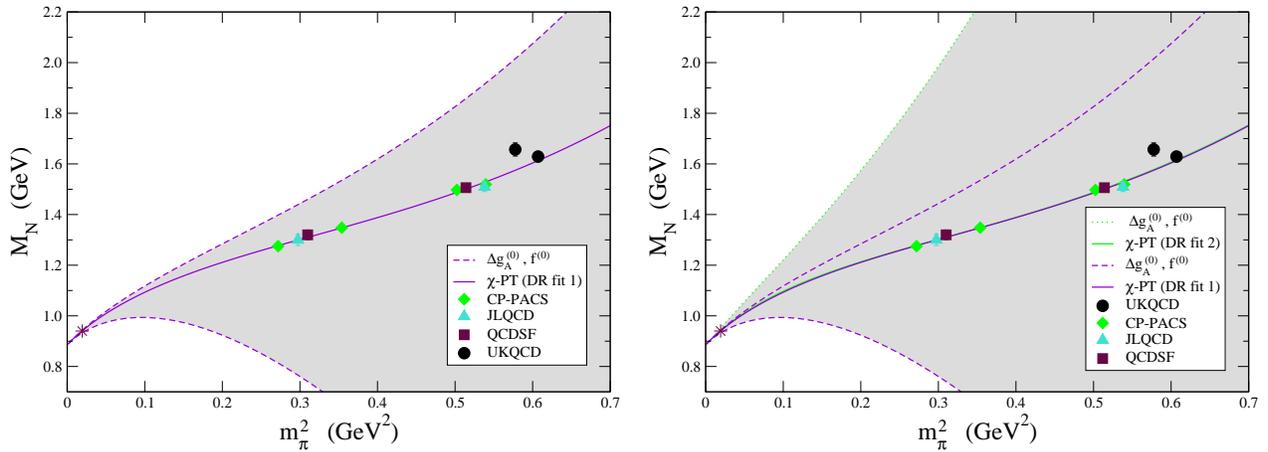

\vskip 0.4in
\centerline{{\epsfxsize=3.2in \epsfbox{swindle2.eps}}\hskip0.36cm{\epsfxsize=3.2in \epsfbox{swindle2A.eps}}} 
\caption{\it Left panel: The nucleon mass in 
baryon $\chi$PT (computed using dim reg with $\overline{\rm MS}$) at ${\cal O}(m_q^2)$
(parameters given in Table~\ref{table:fitjunk} (Fit 1 (DR))) vs. the pion mass squared.
Symbols are unquenched lattice QCD data taken from
Ref.~\protect\cite{Khan:2003cu,UKQCD,CPPACS,JLQCD}. The solid curve is the fit curve
and the dashed curves give maximal variations due to the spread
in chiral limit values of $f$ and $g_A$, as explained in the text.
The gray region corresponds to the error associated with omitted higher orders.
Right panel: Same but with a second fit
(parameters given in Table~\ref{table:fitjunk} (Fit 2 (DR))) with
corresponding spread in the chiral limit values of $f$ and $g_A$.}
\label{fig:swindle2}
\end{figure}

It is interesting to compare the error analysis 
performed using infrared regularization with the same
analysis performed using dimensional regularization (dim reg) with
$\overline{\rm MS}$; as physics must be independent of the regulator,
so must the results of the error analysis~\footnote{Refs.~\cite{Procura,Khan:2003cu} use the nomenclature 
``relativistic $\chi$PT'' to describe infrared regularization; we dislike this appellation
as it may be misinterpreted to suggest that there is physics in the regulator.}. 
The nucleon mass in dim reg with $\overline{\rm MS}$
at ${\cal O}(m_q^2)$ in the chiral expansion is given by~\cite{Steininger:1998ya} 
\begin{eqnarray}
M_N & = & M_0 - 4 c_1 m_\pi^2-\frac{3 g_A^2}{16 \pi f^2} m_\pi^3  
 \nonumber \\ &+&  \left[e_1(\mu)+\frac{3}{32 \pi^2 f^2}
    \left( -\frac{g_A^2}{M_0} + \frac{c_2}{2} \right) 
  - \frac{3}{16 \pi^2 f^2}
       \left( \frac{g_A^2}{M_0} - 8c_1 + c_2 + 4 c_3 \right)
   \ln{\frac{m_\pi}{\mu}} \right] m_\pi^4 \ .
\label{eq:DRformula}
\end{eqnarray}
Again one easily finds a fit to unquenched lattice QCD data over a
wide range of pion masses. In Table~\ref{table:fitjunk} (Fit 1 (DR))
we list a set of parameter values which give rise to the solid curve
in Fig.~\ref{fig:swindle2} (left panel).  Again we consider variations
in the chiral limit values of $f$ and $g_A$ (gray region) and perform
another fit (see Table~\ref{table:fitjunk} (Fit 2 (DR)) and
Fig.~\ref{fig:swindle2} (right panel)).  We find very little
difference in the error analyses for the two regularization schemes.

\begin{figure}[ht]
\vskip 0.4in
\centerline{\epsfxsize=4.2in \epsfbox{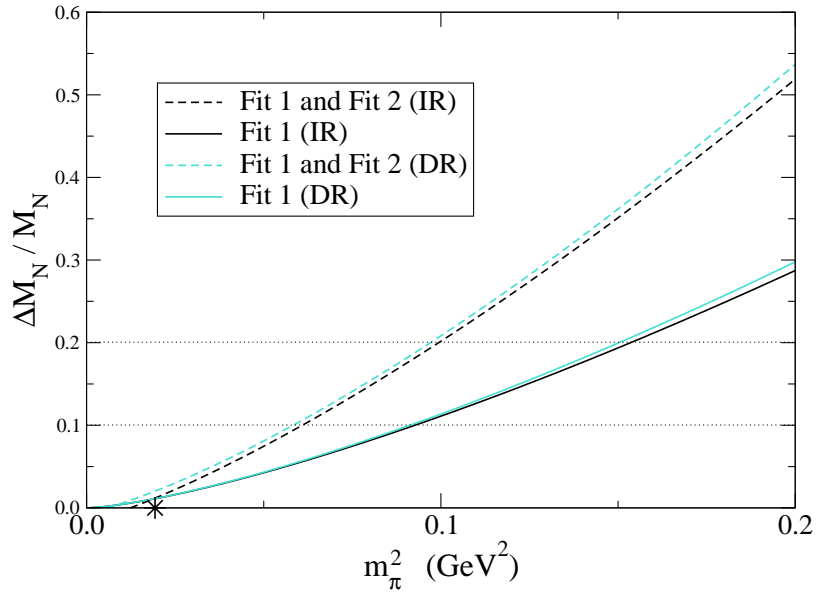}}
\caption{\it Errors in the nucleon mass extrapolation curves vs. the pion mass squared.
The solid lines correspond to the errors abstracted from the left panels of Figs.~\protect\ref{fig:swindle1}
and~\protect\ref{fig:swindle2};
they each arise from the effect of $\Delta g_A^{(0)}$ and $f^{(0)}$ on
a single fit (Fit 1), performed with the physical values of $g_A$ and $f$. 
The dashed lines correspond to the errors abstracted from the right panels of Figs.~\protect\ref{fig:swindle1}
and~\protect\ref{fig:swindle2};
same as above but supplemented with the effect of $\Delta g_A^{(0)}$ and $f^{(0)}$ on a second fit (Fit 2),
performed with select chiral-limit values of $g_A$ and $f$. 
The horizontal dotted lines indicate $10\%$ and $20\%$ errors. The star indicates the physical pion mass.}
\label{fig:errors}
\end{figure}

Our results are summarized in Fig.~\ref{fig:errors}, which plots the
errors in the nucleon mass extrapolation curves taken from
Figs.~\ref{fig:swindle1} and \ref{fig:swindle2}; the curves take into
account the error associated with the gray regions in
Figs.~\ref{fig:swindle1} and \ref{fig:swindle2}. Clearly the chiral
expansion is nicely convergent in the vicinity of the physical pion
mass. In our view the dashed curves (right panels of
Figs.~\ref{fig:swindle1} and~\ref{fig:swindle2}) give a conservative
estimate of the errors. If one is willing to tolerate a $20\%$ error
associated with neglected higher-order terms for an ${\cal O}(m_q^2)$
calculation, then the error analysis would indicate that one is in the
chiral regime for $m_\pi\lsim 300~{\rm MeV}$. If one instead takes the
errors given by the solid curves (left panels of
Figs.~\ref{fig:swindle1} and~\ref{fig:swindle2}), willingness to
tolerate a $20\%$ error would indicate that one is in the chiral
regime for $m_\pi\lsim 400~{\rm MeV}$. It may appear odd that
our estimate of the boundary of the chiral regime is less than the kaon mass, 
which governs the convergence of $SU(3)$ baryon $\chi$PT. We stress that the 
convergence of the chiral expansion is process and flavor dependent;
for instance, some observables in $SU(3)$ baryon $\chi$PT converge well and others do not.

The method used here to estimate errors is, of course, only one method
among many to quantify the reliability of perturbation theory.  For
instance, by comparing various orders in the chiral expansion,
Ref.~\cite{Bernard:2003rp} finds that one is in the chiral regime for
$m_\pi\lsim 500~{\rm MeV}$. An interesting finding of this work is
that the ${\cal O}(m_q^2)$ correction remains a small perturbation on
the ${\cal O}(m_q^{3/2})$ result for $m_\pi<600~{\rm MeV}$.  It is
important to emphasize that error analyses such as that presented in
this note and in Ref.~\cite{Bernard:2003rp} are {\it merely
indicative}. We believe that we have given a conservative estimate of
the errors, as they presently stand.  Naively, one way of weakening
the strength of higher-dimensional operators, and thereby reducing the
error, is to include the $\Delta$ as an explicit degree of freedom in
$\chi$PT~\footnote{As very-few quantities have been computed including
the $\Delta$ at non-trivial orders in baryon $\chi$PT, it is presently unclear whether or
not this is the case.}.  It may also be possible to reduce the
extrapolation error at larger pion masses by computing $M_N$, $g_A$
and $f$ in the same lattice simulation in order to fix the lattice
values of ${d}_{16}$\footnote{ This parameter also provides a
major source of uncertainty in the quark-mass dependence of the
deuteron binding energy~\cite{Beane:2002xf,Epelbaum:2002gb}} and
${l}_{4}$. It is clear from the results presented here that a
definitive determination of the boundary of the chiral regime will not
be possible until these strong-interaction parameters are well determined.
Finally, it is clear that by imposing various
prejudices on our model-independent analysis, one may shrink the gray
regions of Figs.~\ref{fig:swindle1} and~\ref{fig:swindle2} to any
desired size.

In this note we have provided evidence that the quark-mass dependence
---and by association, the finite-size and lattice-spacing dependence--- 
of currently-extant unquenched lattice QCD data for the nucleon mass are not presently
described by a perturbative effective field theory with a controlled
error estimate. Hence these lattice data cannot be reliably extrapolated
to predict nucleon properties. In our view, a symbiotic relationship between lattice QCD and
baryon chiral perturbation theory which will lead to first-principles
predictions of baryon properties with meaningful errors must await
smaller lattice quark masses.

\vskip0.2in


\noindent 
I would like to thank Maarten Golterman, Thomas Hemmert, Ulf Mei\ss ner and Martin Savage for helpful
comments/discussions. This work was partly supported by DOE contract
DE-AC05-84ER40150, under which the Southeastern Universities Research
Association (SURA) operates the Thomas Jefferson National Accelerator
Facility.

\vfill\eject

\end{document}